\documentclass[showpacs,amsmath,amssymb,nofootinbib,prd]{revtex4}
\usepackage{graphicx}
\usepackage{enumerate}
\usepackage{amsmath}
\usepackage{amssymb}
\usepackage{amsfonts}
\usepackage{color}
\usepackage{pstricks}
\usepackage{color}

\begin{document}
\title{Rare top decay $t \rightarrow c \gamma$ with flavor changing neutral scalar interactions in two Higgs doublet model}
\author{R. Gait\'an}
\email{rgaitan@unam.mx} \affiliation{Departamento de F\'isica, FES-Cuautitl\'an, UNAM, C.P. 54770, Estado de M\'exico, M\'exico}
\author{E. A. Garc\'es}
\email{egarces@fisica.unam.mx}\affiliation{Instituto de F\'isica, Universidad Nacional Aut\'onoma de M\'exico, Ciudad de M\'exico, 01000, M\'exico.}
\author{R. Martinez}
\email{remartinezm@unal.edu.co} \affiliation{Departamento de F\'isica, Universidad Nacional de Colombia, Bogot\'a D.C., Colombia}
\author{J.H. Montes de Oca}
\email{josehalim@gmail.com}\affiliation{Departamento de F\'isica, FES-Cuautitl\'an, UNAM, C.P. 54770, Estado de M\'exico, M\'exico}
\begin{abstract}
{Models beyond the Standard Model with extra scalars have been highly motivated by the recent discovery of a Higgs boson. The Two Higgs Doublet Model Type III considers the most general case for the scalar potential, allowing mixing between neutral CP-even and CP-odd scalar fields. This work presents the results of the study on the $t\rightarrow c\gamma$ decay at one loop level if neutral flavor changing is generated by top-charm-Higgs coupling given by the Yukawa matrix. For instance, a value for the branching ratio $Br(t\rightarrow c\gamma)\sim 10^{-6}$ for $\tan\beta=2.5$ and general neutral Higgs mixing parameters, $1.16\leq\alpha_1\leq1.5$, $-0.48\leq\alpha_2\leq-0.1$. The number of events for the $t\rightarrow c \gamma$ decay with an integrated luminosity of 300 $fb^{-1}$ is estimated as $10\lesssim N_{Eff}\lesssim 100$ for the parameters of the model constrained by experimental data.}
\end{abstract}
\pacs{14.65.Ha, 14.80.Bn, 12.60.-i}
%
\maketitle
%
%
\section{Introduction}
\label{sec1}

The observation of the scalar-like Higgs boson with a mass of 126 GeV at the Large Hadron Collider (LHC) \cite{Aad:2012tfa,Chatrchyan:2012xdj} has motivated the study of extended models with multiple scalar multiplets. The mass hierarchy between the up-type and down-type quarks suggests the consideration of models with two complex $SU(2)_L$ doublet scalar fields, referred to as Two Higgs Doublet Models (THDM).
There are two versions of THDM, labeled as type I and type II, with invariance under a $Z_2$ discrete symmetry which ensures CP conservation in the scalar sector \cite{hunter}. In the first case, all quarks acquire mass through one doublet~\cite{Haber:1978jt,Hall:1981bc} whereas in type II~\cite{Donoghue:1978cj} one doublet gives mass to the up-type quarks while the other doublet gives mass to the down-type quarks.

In the so called type III both doublets simultaneously give masses to all quark types, which will hence be referred as Model III \cite{Atwood:1996vj}. In any type of THDM five physical Higgs particles are predicted, three of them are neutral with CP-even or CP-odd states and a charged pair. An important feature in Model III is the mixing between the CP-even and CP-odd states for neutral scalar fields given by the mixing parameters $\alpha_1, \alpha_2$ and $\alpha_3$~\cite{Barroso:2012wz,Lavoura:1994fv,Fontes:2015mea}. Current measurements in LHC imply that the $126$ GeV scalar particle is in good agreement with the Higgs boson being CP even~\cite{ATLAS:2013nma,Khachatryan:2014kca}.

Model III without $Z_2$ discrete symmetry is a general version that generates Flavor Changing Neutral Scalar Interactions (FCNSI) in Higgs-fermion Yukawa couplings and CP violation in the Higgs potential \cite{Cheng:1987rs,Crivellin:2013wna,Fritzsch:1977za,Fritzsch:1977vd,Fritzsch:1979zq}. One motivation to look for new sources of CP violation beyond the SM is the matter-antimatter problem \cite{Shu:2013uua,Morrissey:2012db} as well as the fermion electric dipole moments \cite{Inoue:2014,Carcamo:2006dp,Cvetic:1997zd,Cvetic:1998uw}. On the side of the FCNSI, a motivation arises from the study of the Flavor Changing Neutral Current (FCNC) processes, which are extremely suppressed in the Standard Model (SM), for instance $\textrm{Br}(t\rightarrow q+ x)\approx 10^{-17}-10^{-12}$ with $q=c,\,u$ and $x=\gamma,\,\,Z,\,\,g,\,H$~\cite{ Eilam:1990zc, AguilarSaavedra:2004wm, Grzadkowski:1990sm, Mele:1999zx, Mele:1998ag, Gabrielli:2011yn, Eilam:1990zc, AguilarSaavedra:2002ns, Larios:2006pb, DiazCruz:1989ub}. In particular we are interested in the $t\rightarrow c \gamma$ rare decay. The LHC excludes the ranges of $\textrm{Br}(t\rightarrow c \gamma)> 5.9\times10^{-3}$, meanwhile in future results it is expected to set an upper bound of order $ 10^{-5}$ \cite{Agashe:2014kda}.

In ~\cite{Luke:1993cy} is estimated a value for $\textrm{Br}\left( t\rightarrow c \gamma\right)\sim 10^{-8}$ with charged Higgs mass $m_{H^{\pm}} \sim $ 200 GeV as well as small values of the $\beta$ mixing parameter, $\tan\beta=0.1$. A detailed study in the framework of Model III with FCNC shows more feasible values for branching ratio in the range $10^{-12}<\textrm{Br}\left( t\rightarrow c \gamma\right) < 10^{-7}$ with the masses of the scalars between 200 GeV and 800 GeV~\cite{Atwood:1995ud,Atwood:1996vw,Atwood:1995ej,Atwood:1996vj,Arhrib:2005nx}. For the different THDM types, the $\textrm{Br}\left( t\rightarrow c \gamma\right)$ is enhanced for specific regions of scalar masses and mixing parameters \cite{AguilarSaavedra:2002ns,Branco:2011iw,AguilarSaavedra:2002kr}. 

The rare top decay has been analyzed in extended models other than THDM, for instance \cite{Han:2009zm,HongSheng:2007ve,GonzalezSprinberg:2007zz,Cao:2007dk,Han:2003qe}. In a previous work~\cite{Diaz:2001vj,Gaitan-Lozano:2014nka}, it was shown that $\textrm{Br}\left( t\rightarrow c \gamma\right)$ is sensitive to $\tan\beta$ in the framework of Model III, obtaining $\textrm{Br}\left( t\rightarrow c \gamma\right)\sim 1\times10^{-6}$ for $8\leq\tan\beta\leq15$. The rare top-quark decays at one loop with FCNC coming from additional fermions and gauge bosons has been studied in several extensions of the SM such as MSSM, Left-Right symmetry Models, top color assisted technicolor, little Higgs and two Higgs doublets with four generations of quarks \cite{Atwood:1995ud,Atwood:1996vw,Atwood:1995ej,Atwood:1996vj,Arhrib:2005nx,Luke:1993cy,DiazCruz:1989ub,Dedes:2014asa}. FCNC and CPV between quarks and scalars can also contribute to interactions with  rare top decay \cite{Accomando:2006ga,Niezurawski:2004ui}.

The content of this paper is as follows. The next section introduces the model and the interactions between quarks and neutral Higgs bosons. In section \ref{sec3}, we calculate $Br (t \rightarrow c \gamma) $ in the framework of the Model III with FCNSI including CP violation in the scalar sector, in section \ref{sec4} we present the restrictions to the parameters involved in the rare top decay. We present the results of our analysis  
in section \ref{sec5}. Finally, the conclusion is stated in section \ref{sec6}.
%
%
\section{Flavor Changing Neutral Scalar Interactions}
\label{sec2}
Given $\Phi_1$ and $\Phi_2$ two complex $SU(2)_L$ doublet scalar fields with hypercharge-one, the most general gauge invariant and renormalizable Higgs scalar potential is
\cite{Haber:1993an}
\begin{eqnarray}
V &=&m_{11}^{2}\Phi _{1}^{+}\Phi _{1}+m_{22}^{2}\Phi _{2}^{+}\Phi _{2}-\left[
m_{12}^{2}\Phi _{1}^{+}\Phi _{2}+h.c.\right] +\frac{1}{2}\lambda _{1}\left(
\Phi _{1}^{+}\Phi _{1}\right) ^{2}  
+ \frac{1}{2}\lambda _{2}\left( \Phi _{2}^{+}\Phi _{2}\right) ^{2} \nonumber \\ 
&& + \lambda_{3}\left( \Phi _{1}^{+}\Phi _{1}\right) \left( \Phi _{2}^{+}\Phi_{2}\right) 
  +\lambda _{4}\left( \Phi _{1}^{+}\Phi _{2}\right)  \left( \Phi_{2}^{+}\Phi _{1}\right)  \nonumber \\ 
  &&+\left[ \frac{1}{2}\lambda _{5}\left( \Phi _{1}^{+}\Phi _{2}\right)
^{2}+\lambda _{6}\left( \Phi _{1}^{+}\Phi _{1}\right) \left( \Phi
_{1}^{+}\Phi _{2}\right) +\lambda _{7}\left( \Phi _{2}^{+}\Phi _{2}\right)
\left( \Phi _{1}^{+}\Phi _{2}\right)  + h.c. \right] ,
\end{eqnarray}
where $m_{11}^2$, $m_{22}^2$ and $\lambda_1$, $\lambda_2$, $\lambda_3$, $\lambda_4$ are real parameters and $m_{12}^2$, $\lambda_5$, $\lambda_6$, $\lambda_7$ can be complex parameters. The most general $U(1)_{EM}$-conserving vacuum expectation values (VEV) are
\begin{equation}
\langle \Phi_1 \rangle= \frac{1}{\sqrt{2}}\left(
\begin{array}{c}
0 \\
v_1 \\
\end{array}
\right),
\label{vev1}
\end{equation}
\begin{equation}
\langle \Phi_2 \rangle= \frac{1}{\sqrt{2}}\left(
\begin{array}{c}
0 \\
v_2 e^{i\xi}\\
\end{array}
\right),
\label{vev2}
\end{equation}
where $v_1$ and $v_2$ are real and non-negative, $0 \leq |\xi| \leq \pi$, and $v^2 \equiv v_1^2 + v_2^2 = \frac{4 M_W^2}{g^2} = \left (246 ~\textrm{GeV} \right)^2$. Without loss of generality, the phase in the Eq. (\ref{vev1}) was eliminated through the $U(1)_Y$ global invariance, leaving the $\xi$ phase in the VEV of Eq. (\ref{vev2}). This $\xi$ phase is a source of spontaneous CP violation which can be absorbed by redefining the free parameters \cite{Ginzburg:2004vp}.

The neutral components of the scalar Higgs doublets in the interaction basis are $\frac{1}{\sqrt{2}}\left( v_a + \eta_a +i \chi_a\right)$, where $a=1,2$. As a result of the explicit CP symmetry breaking, a mixing matrix $R$ relates the mass eigenstates $h_i$ with the $\eta_i$ as follows
\begin{equation}
h_{i}=\sum_{j=1}^{3}R_{ij}\eta _{j},
\label{h-Rn}
\end{equation}
where the state orthogonal to the Goldstone boson associated to $Z$ boson is $\eta_3=-\chi_1 $ $\sin\beta+\chi_2\cos\beta$ and $R$ is parametrized as \cite{ElKaffas:2006gdt}:
\begin{equation}
R=\left(
\begin{array}{ccc}
c_{1}c_{2} & s_{1}c_{2} & s_{2} \\
-\left( c_{1}s_{2}s_{3}+s_{1}c_{3}\right)
& c_{1}c_{3}-s_{1}s_{2}s_{3} & c_{2}s_{3} \\
-c_{1}s_{2}c_{3}+s_{1}c_{3} & -\left(
c_{1}s_{1}+s_{1}s_{2}c_{3}\right)  &
c_{2}c _{3}
\end{array}
\right),
\label{r_matrix}
\end{equation}
with $c_i=\cos\alpha_i$, $s_i=\sin\alpha_i$ for $-\frac{\pi}{2}\leq\alpha_{1,2}\leq\frac{\pi}{2}$ and $0\leq\alpha_3\leq\frac{\pi}{2}$. The neutral Higgs bosons $h_i$ satisfy the mass relation $m_{h_1}\leq m_{h_2}\leq m_{h_3}$~\cite{Basso:2012st,Arhrib:2010ju,Krawczyk:2013jta,Chen:2015gaa}. In the CP conserving case $\eta_{1}$ and $\eta_{2}$ are CP-even and mixed in a $2\times2$ matrix while $\eta_3$ is CP-odd without mixing with $\eta_{1}$ and $\eta_{2}$. However, due to the CP-symmetry breaking in the general case, the neutral Higgs bosons $h_{1,2,3}$ do not have well defined CP states.
The most general structure for the Yukawa couplings among fermions and scalar is
\begin{equation}
\mathcal{L}_{Yukawa}=\sum_{i,j=1}^{3}\sum_{a=1}^{2}\left( \overline{q}
_{Li}^{0}Y_{aij}^{0u}\widetilde{\Phi }_{a}u_{Rj}^{0}+\overline{q}
_{Li}^{0}Y_{aij}^{0d}\Phi _{a}d_{Rj}^{0}+\overline{l}_{Li}^{0}Y_{aij}^{0l}
\Phi _{a}e_{Rj}^{0}+h.c.\right) ,  \label{yukawa}
\end{equation}
where $Y_{a}^{u,d,l}$ are the $3\times 3$ Yukawa matrices. $q_{L}$ and $l_{L}$
denote the left handed fermion doublets under $SU(2)_L$, while $u_{R}$, $d_{R}$, $l_{R}$ correspond to the right handed singlets. The zero superscript in fermion fields stands for the interaction basis. After getting a correct spontaneous symmetry breaking by the VEV using Eq.(\ref{vev1}) and Eq.(\ref{vev2}), the mass matrices become
\begin{equation}
M^{u,d,l}=\sum_{a=1}^{2}\frac{v_{a}}{\sqrt{2}}Y_{a}^{u,d,l},  \label{mass}
\end{equation}
where $Y_a^{f}=V_L^f Y_a^{0f}\left(V_R^{f}\right)^\dag$, for $f=u,d,l$. The $V_{L,R}^f$ matrices are used to diagonalize the fermion mass matrices and relate the physical and interaction states. Note that in Model III the diagonalization of mass matrices does not imply the diagonalization of the Yukawa matrices, as it happens in the THDM type I or II. An important consequence of non-diagonal Yukawa matrices in physical states is the presence of FCNSI between neutral Higgs bosons and fermions.

The focus is on the up-type quark Yukawa interactions that contain the Feynman rules for the rare top decay. Replacing from Eq.(\ref{h-Rn}) and Eq.(\ref{mass}) in the Yukawa Lagrangian of Eq.(\ref{yukawa}), the interactions between neutral Higgs bosons and fermions can be written as interactions of the THDM with CP conserving (type I or II) plus additional contributions, which arise from any of the $Y_{1,2}$ Yukawa matrices. The relation among the mass matrix $M^F$ and  the Yukawa matrices $Y_{1,2}^F$, for $F=u,d,l$, is used to write the Yukawa Lagrangian, Eq.(\ref{yukawa}), as a function only of one Yukawa matrix, $Y_1^F$ or $Y_2^F$. We choose to write the interactions as a function of the Yukawa matrix $Y_2$, that is, $Y_1^F= \frac{\sqrt{2}}{v_1}M^F-\frac{v_2}{v_1}Y_2^F$ is replaced in Eq.(\ref{yukawa}). From now on, in order to simplify the notation, the subscript 2 in the Yukawa couplings will be omitted. The interactions between quarks and Higgs bosons in the mass eigenstates are explicitly written as
\begin{eqnarray}
\mathcal{L} &=&\frac{1}{v\cos \beta }\sum_{ijk}\bar{u}_{i}M_{ij}^{u}\left(
A_{k}P_{L}+A_{k}^{\ast }P_{R}\right) u_{j}h_{k}  
+\frac{1}{v\cos \beta }\sum_{ijk}\bar{d}_{j}M_{ij}^{d}\left( A_{k}^{\ast
}P_{L}+A_{k}P_{R}\right) d_{j}h_{k}  \nonumber \\
&&+\frac{1}{\cos \beta }\sum_{ijk}\bar{u}_{i}Y_{ij}^{u}\left(
B_{k}P_{L}+B_{k}^{\ast }P_{R}\right) u_{j}h_{k} 
+\frac{1}{\cos \beta }\sum_{ijk}\bar{d}_{i}Y_{ij}^{d}\left( B_{k}^{\ast
}P_{L}+B_{k}P_{R}\right) d_{j}h_{k}  \nonumber \\
&&+\left[ \frac{\sqrt{2}}{\cos \beta }\sum_{ij}\bar{u}_{i}\left(
(KY^{d})_{ij}P_{R}-(Y^{u}K)_{ij}P_{L}\right) d_{j}H^{+}\right.   \nonumber \\
&&+\frac{\sqrt{2}}{v}\tan \beta \sum_{ij}\bar{u}_{i}\left( -\left(
KM^{d}\right) _{ij}P_{R}+\left( M^{u}K\right) _{ij}P_{L}\right) d_{j}H^{+}
\nonumber \\
&&\left. +\frac{\sqrt{2}}{v}\bar{u}_{i}\sum_{ij}\left( \left( M^{d}K\right)
_{ij}P_{R}-\left( M^{u}K\right) _{ij}P_{L}\right) d_{j}G_{W}^{+}+h.c.\right]
,  \label{yukawa_quarks}
\end{eqnarray}
where we define
\begin{eqnarray}
A_k&=& R_{k1} - i R_{k3}\sin\beta ,\nonumber \\
B_k&=& R_{k2} \cos\beta - R_{k1} \sin\beta + i R_{k3}.
\label{ak}
\end{eqnarray}
The fermion spinors are denoted as $(u_1,\,u_2,\,u_3)=(u,\,c,\,t )$, where the indexes $i,\,j=1,2,3$ denote the family generations in Eq. (\ref{yukawa_quarks}), while $k=1,2,3$ is used for the neutral Higgs bosons and $P_{R,L}=\frac{1}{2}\left(1\pm \gamma_5\right)$. Note that a CP conserving case is obtained only if two neutral Higgs bosons are mixed with well-defined CP states, for instance $\alpha_2=\alpha_3=0$ is the usual limit.
%
%
\section{Rare top decay $t\to c\gamma$}
\label{sec3}
The expression for the $t\rightarrow c\gamma$ decay amplitude is a magnetic transition written as
\begin{equation}
\mathcal{M} = \bar{u}\left( p^\prime \right)\left[ F_1 \sigma_{\mu\nu}+F_2 \sigma_{\mu\nu}\gamma_5\right] q^\nu u\left( p \right) \epsilon^\mu\left( q \right),
\label{ampli}
\end{equation}
where $p^\prime=p-q$, $\epsilon^\mu\left( q \right) $ is the photon polarization; when the photon is on-shell, $q^2=0$, and $\epsilon^\mu\left( q \right)q_\mu=0$. The invariant amplitudes $F_{1,2}$ are obtained in terms of the model parameters as shows Eq.(\ref{width}). Eq.(\ref{ampli}) corresponds to a five-dimension operator, and then the on-shell $t\rightarrow c\gamma$ amplitude must be represented by a set of loop diagrams.
%
%
\begin{figure}
\begin{minipage}{\columnwidth}
\centering
\includegraphics[scale=0.5]{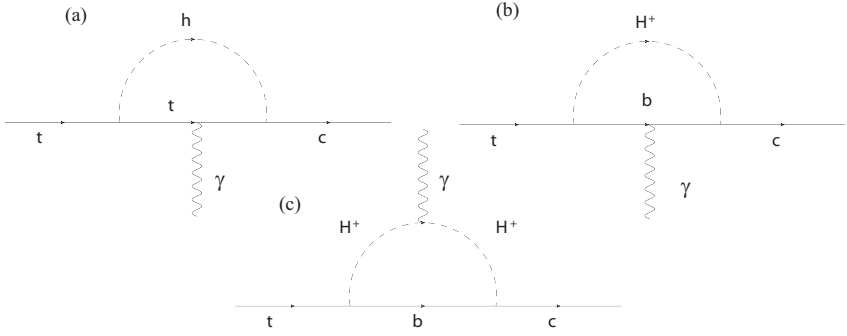}
\end{minipage}
\caption{\label{dia1} One loop Feynman diagram with a Higgs boson in the internal line, (a) flavor changing neutral scalar contribution, (b) and (c) charged contributions.}
\end{figure}
Fig.(\ref{dia1}) shows the dominant contributions for the rare top decay $t\rightarrow c \gamma$ at one loop coming from neutral and charged Higgs bosons. The charged contributions, see Fig.1(b) and Fig.1(c), are suppressed by the bottom quark mass compared to the top quark mass in the neutral Higgs contribution. In order to study the effects of FCNSI we analyze only the dominant contribution, see Fig.1(a). In order to obtain the partial width of the $t \rightarrow c \gamma$ decay in Model III we apply the method previously used in \cite{DiazCruz:1989ub}. Integrating over the internal momentum, the partial width is 
\begin{equation}
\Gamma \left( t\rightarrow c\gamma \right) =\frac{\alpha G_{F}m_{t}^{3}}{%
192\pi ^{4}\cos ^{4}\beta }\left\vert Y_{ct}^{u}\right\vert ^{2}\sum_{k}%
\left\vert f_{1}\left( \widehat{m}_{k}\right) A_{k}^{\ast
}B_{k}+f_{2}\left( \widehat{m}_{k}\right) A_{k}B_{k}^{\ast
}\right\vert ^{2}, 
\label{width}
\end{equation}
where $G_F^{-1}=\sqrt{2}v^2$, $v = 246$ GeV, $\alpha\approx1/128$ at electroweak scale and the functions $f_{1,2}$ are defined as
\begin{equation}
f_{1}\left( \widehat{m}_{k}\right) =\int_{0}^{1}dx\int_{0}^{1-x}dy\frac{
x\left( x+y-1\right) }{x^{2}+xy-\left( 2-\widehat{m}_{k}^{2}\right) x+1},
\end{equation}%
\begin{equation}
f_{2}\left( \widehat{m}_{k}\right) =\int_{0}^{1}dx\int_{0}^{1-x}dy\frac{
\left( x-1\right) }{x^{2}+xy-\left( 2-\widehat{m}_{k}^{2}\right) x+1},
\end{equation}
with $\widehat{m}_{i}=\frac{m_{h_i}}{m_{t}}$ for $i=1,2,3$. The branching ratio can be approximated as
\begin{equation}
\textrm{Br}\left( t\rightarrow c\gamma \right)\approx\frac{\Gamma \left( t\rightarrow
c\gamma \right) }{\Gamma _{\textrm{top}}},
\label{br}
\end{equation}
where $\Gamma_{\textrm{top}}$ at NLO is given by \cite{Agashe:2014kda}
\begin{equation}
\Gamma_{\textrm{top}}=\frac{G_f m_t^3}{8\pi\sqrt{2}} \left(1-\frac{M_W^2}{m_t^2}\right)^2 \left(1-2\frac{M_W^2}{m_t^2}\right)\left[1-\frac{2\alpha_s} {3\pi}  \left( \frac{2\pi^2}{3}-\frac{5}{2} \right)\right].
\end{equation}
%
%
\section{Constraints on rare top decay parameters}
\label{sec4}
\begin{figure} 
\includegraphics[scale=0.6]{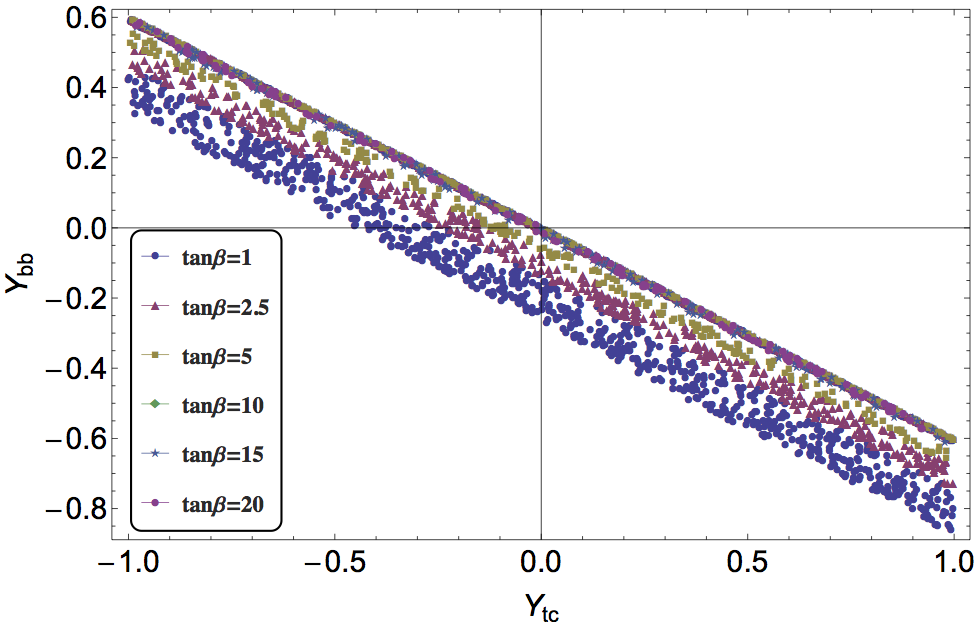}
\caption{Allowed values for the Yukawa couplings, scatter plot with points compatible with the experimental value of the $BR (B\to X_s\gamma)$ and $300$ GeV $\leqslant m_H^{\pm} \leqslant 600$ GeV and $\tan\beta=$ 1, 2.5, 5, 10 and 15 .}
\label{fig:fig_bsg}
\end{figure}

We note that Eq.(\ref{width}) contains free parameters of the THDM, such as the masses of the neutral Higgs bosons, the mixing angles $\alpha_i$, $\beta$ and Yukawa couplings. In order to set allowed values for free parameters we first review the possible constraints that $b\rightarrow s \gamma$ decay can impose on the $Y_{tc}$ coupling.
Following references \cite{Degrassi:2000qf,Misiak:2006zs,Lunghi:2006hc,Gomez:2006uv,Barenboim:2013bla}, the branching ratio of the $b\rightarrow s \gamma$ decay is a function of the Wilson
coefficients and it can be written as:
\begin{eqnarray}
\textrm{Br}(B \rightarrow X_{s}\gamma )&\approx& a+a_{77}\delta
C_{7}^{2}+a_{88}\delta C_{8}^{2}+Re\left( a_{7}\delta C_{7}\right) +Re\left(
a_{8}\delta C_{8}\right) + Re\left( a_{78}\delta C_{7}\delta C_{8}^{\ast }\right) ,
\end{eqnarray}
with $a\approx 3.0\times 10^{-4}$, $a_{77} \approx 4.7\times 10^{-4}$, $a_{88} \approx 0.8\times10^{-4}$, $a_7 \approx (-7.2+0.6i)\times 10^{-4}$,
$a_8 \approx (-2.2 - 0.6 i) \times 10^{-4}$ and $a_{78} \approx (2.5 - 0.9 i) \times 10^{-4}$. The main contributions due to Wilson coefficients, beyond the W-boson contribution, are given by charged Higgs and flavor changing (FC) Yukawa couplings, $\delta C_{7,8}=C^{H^{\pm}_{7,8}} +C^{H,FC}_{7,8}$.
The charged-Higgs contribution is
\begin{equation}
C^{H^{\pm}_{7,8}} =\frac{1}{3\tan^2\beta} f_{7,8}^{(1)}(y_t) + f_{7,8}^{(2)}(y_t),
\label{eq_h}
\end{equation}
while the FC contribution is 
\begin{eqnarray}
C_{7,8}^{H,FC} &=&\frac{2M_{W}}{gm_{t}K_{ts}\cos \beta }%
(Y^{u}K)_{ts}f_{7,8}^{(2)}(y_{t}) 
+ \frac{2M_{W}}{gm_{b}K_{tb}\cos \beta }(KY^{d})_{tb}f_{7,8}^{(2)}(y_{t})
\label{eq_chfc}
\end{eqnarray}
with $y_t=m^2_t/M^2_{H}$ and the explicit relations $f_{7,8}^{(1),(2)}(x)$ can be found in Ref. \cite{Degrassi:2000qf,Misiak:2006zs,Lunghi:2006hc,Gomez:2006uv,Barenboim:2013bla}.
Using the hierarchy of the Kobayashi-Maskawa matrix $(K)$ we have the following approximations $(Y^uK)_{ts}\approx Y_{tc}K_{cs}$ and $(KY^d)_{tb}\approx K_{tb}Y_{bb}$. In order to have a bound to the $Y_{tc}$ FC Yukawa coefficient, it was considered that $(KY^d)_{tb}$ gives the most important contribution.
The limits on the $B\to X_s\gamma$ decay come from BaBar, Belle and CLEO \cite{Chen:2001fja,Abe:2001hk,Lees:2012wg,Lees:2012wg,Lees:2012ufa,Aubert:2007my}. The current world average for $E> 1.6$~GeV, given by HFAG~\cite{Amhis:2014hma}, is
\begin{eqnarray}
\textrm{Br} (B\to X_s\gamma) = (3.43 \pm 0.21 \pm 0.07) \times 10^{-4} .
\label{br_bsg}
\end{eqnarray}
This result provides an important constraint on the ($Y_{tc}$, $Y_{bb}$) space, Fig.(\ref{fig:fig_bsg}), with $m_{H^{\pm}}=500$ GeV and $0<\tan\beta<20$.
%
%
%
%
\begin{figure} 
\includegraphics[scale=0.5]{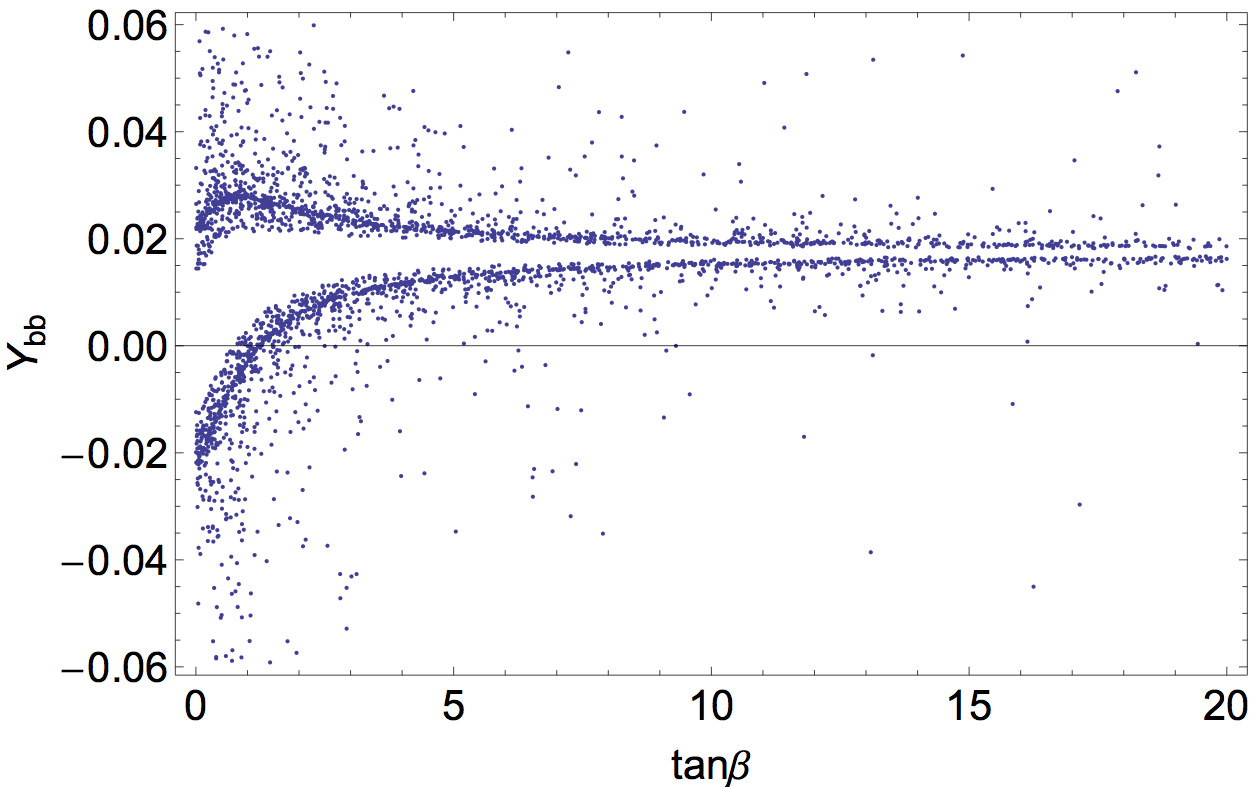}
\caption{Allowed values for the Yukawa couplings $Y_{bb}$, scatter plot with compatible points with the experimental value of the $BR (H\to b\bar{b})$ and $\pi/2\leqslant \alpha_{1,2} \leqslant \pi/2 $.}
\label{ybb_beta}
\end{figure}
%
%
\begin{figure} 
\includegraphics[scale=0.7]{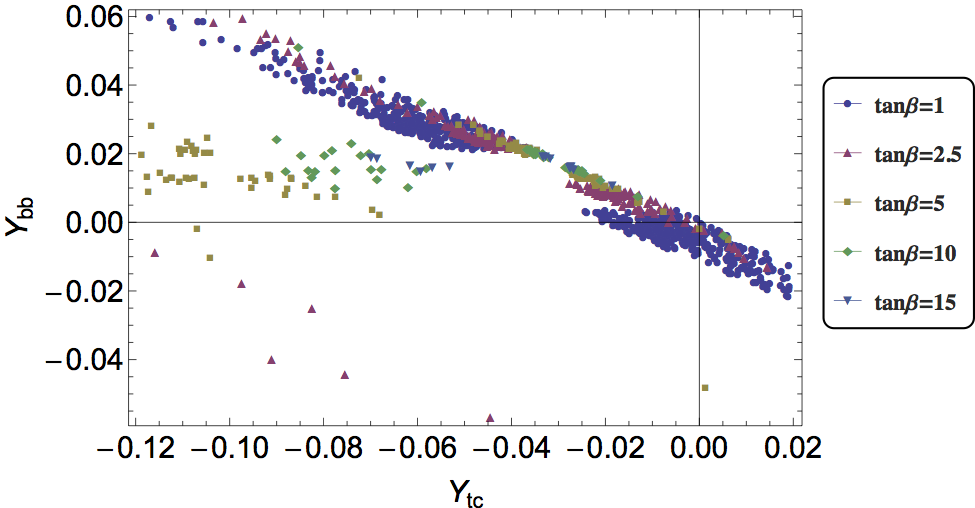}
\caption{Allowed values for the Yukawa couplings, scatter plot with compatible points with the experimental values for  $Br(H\to b\bar{b})$ and  $Br(B\to X_s\gamma)$.}
\label{fig:}
\end{figure}

The second constraint considered is based in the branching ratio of the SM Higgs boson decay to bottom quark pairs, which has a reported value of  $Br\left(H\rightarrow b \bar{b}\right) = 5.77\times 10^{-1}{}^{+3.2\%}_{-3.3\%}$ \cite{Agashe:2014kda}. The width decay in the THDM for $h_1\rightarrow b \bar{b}$  is given by
\begin{equation}
\Gamma_{h_1\rightarrow b \bar{b}}=\frac{N_{c}m_{h1}}{8\pi}\left(1-4\frac{m_b^2}{m_{h1}^2}\right)^{\frac{1}{2}} \left[C^2\left(1-4\frac{m_b^2}{m_{h1}^2}\right)+D^2 \right],
\label{w_hbb}
\end{equation}
where 
\begin{equation}
\label{c2}
C^2 = \left[ \frac{m_b}{v\cos\beta}R_{11}+\frac{Y_{bb}}{\cos\beta}\left( R_{12}\cos\beta-R_{11}\sin\beta \right) \right]^2
\end{equation}
and
\begin{equation}
\label{d2}
D^2=\left[- \frac{m_b}{v\cot\beta}R_{13}+ \frac{Y_{bb}}{\cos\beta}R_{13}\right]^2.
\end{equation}
Note that the matrix elements $R_{11}$, $R_{12}$ and $R_{13}$ are independent of the mixing parameter $\alpha_3$. Figure (\ref{ybb_beta}) shows the behavior of $Y_{bb}$ as function of $\tan\beta$ for random values of $\alpha_{1,2}$. After that previous constrains are imposed, the allowed values for Yukawa couplings are $-0.02\leqslant Y_{bb} \leqslant 0.06$ and $-0.12\leqslant Y_{tc}\leqslant 0.02 $ for $1\leqslant \tan\beta\leqslant 15$, see Figure (\ref{fig:}). The non-diagonal elements of the Yukawa matrix responsible of the FCNSI, shown in Eq.(\ref{yukawa_quarks}), must be suppressed \cite{Glashow:1976nt}.

%
%
\section{Results}
\label{sec5}

Focusing on the rest of the parameters, note that the masses of the $h_i$ neutral Higgs bosons are set so that the mass of the lightest Higgs boson $h_1$ is equal to the mass value of the observed scalar reported by ATLAS and CMS, $m_{h_1}\approx 126$ GeV \cite{Aad:2012tfa,Chatrchyan:2012xdj}.  Contributions to Eq.(\ref{br}) from $h_2$ and $h_3$ are negligible for masses $m_{h_2},\,m_{h_3}>600$ GeV. 

\begin{figure} 
\includegraphics[scale=0.8]{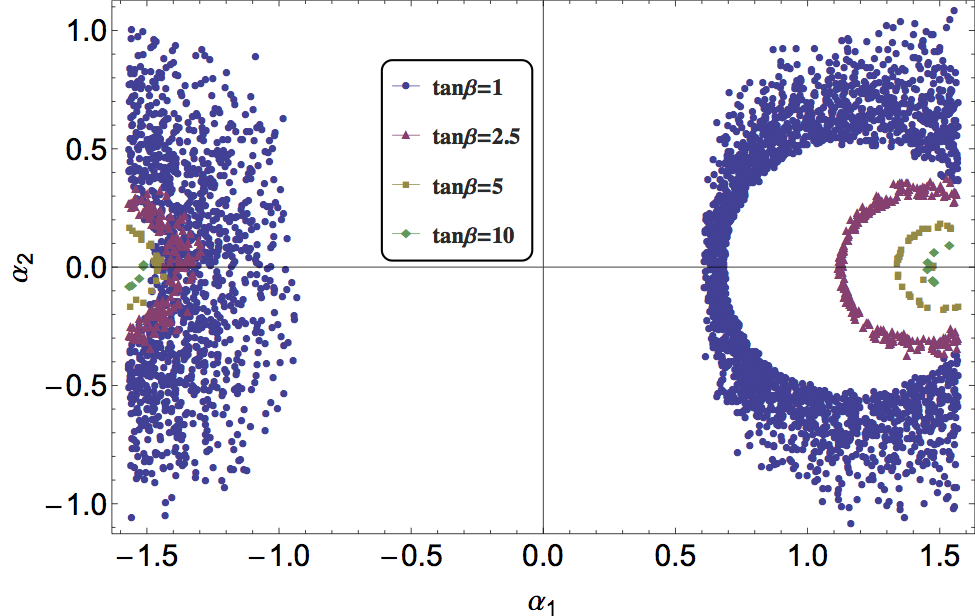}
\caption{Allowed values for the mixing parameter $\alpha_{1,2}$, scatter plot with points compatible with the experimental values for  $Br(H\to b\bar{b})$, $Br(B\to X_s\gamma)$ and  $Br(t\to c \gamma)<5.9\times10^{-3}$ , for fixed values of $\tan\beta=1,2.5, 5, 10.$ and $Y_{tc}=0.01$}
\label{a1a2y1}
\end{figure}

\begin{figure} 
\includegraphics[scale=0.8]{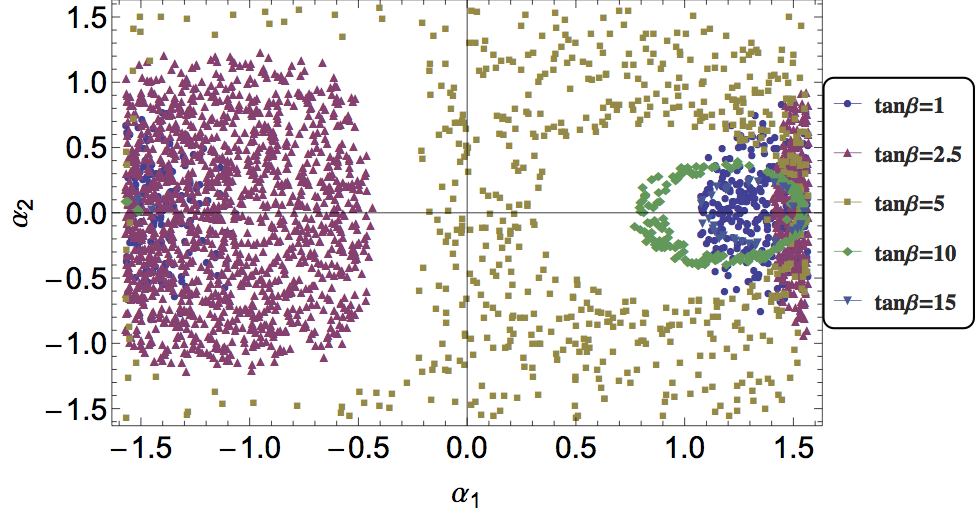}
\caption{Allowed values for the mixing parameters $\alpha_{1,2}$, scatter plot with points compatible with the experimental values for  $Br(H\to b\bar{b})$, $Br(B\to X_s\gamma)$ and  $Br(t\to c \gamma)<5.9\times10^{-3}$ , for fixed values of $\tan\beta=1,2.5, 5, 10.$ and $Y_{tc}=-0.4$}
\label{a1a2y2}
\end{figure}

Also, note that the contribution from $h_1$ is independent of the mixing parameter $\alpha_3$, see the first row in matrix Eq.(\ref{r_matrix}). Therefore, the set of free parameters considered in the partial width Eq.(\ref{width}) is reduced only to the mixing angles $\left\{\alpha_1,\,\alpha_2,\,\beta\right\}$. Figures (\ref{a1a2y1}) and (\ref{a1a2y2}) show the allowed values for mixing parameters $\alpha_1$ and $\alpha_2$ when the current  limit for the $Br(t\rightarrow c \gamma)<5.9\times10^{-3}$ is considered \cite{Agashe:2014kda}. Based in  Fig.(\ref{fig:}) the Yukawa coupling $Y_{tc}$ was fixed with the two representative values $Y_{tc}= -0.04, 0.01$. 
%
%
%
%

In order to analyze the $Br(t\rightarrow c \gamma)$ we consider the allowed regions for the mixing parameters $\alpha_1$ and $\alpha_2$ previously fixed in \cite{Gaitan:2013yfa}.
%
%
\begin{figure}[h]
\includegraphics[scale=0.8]{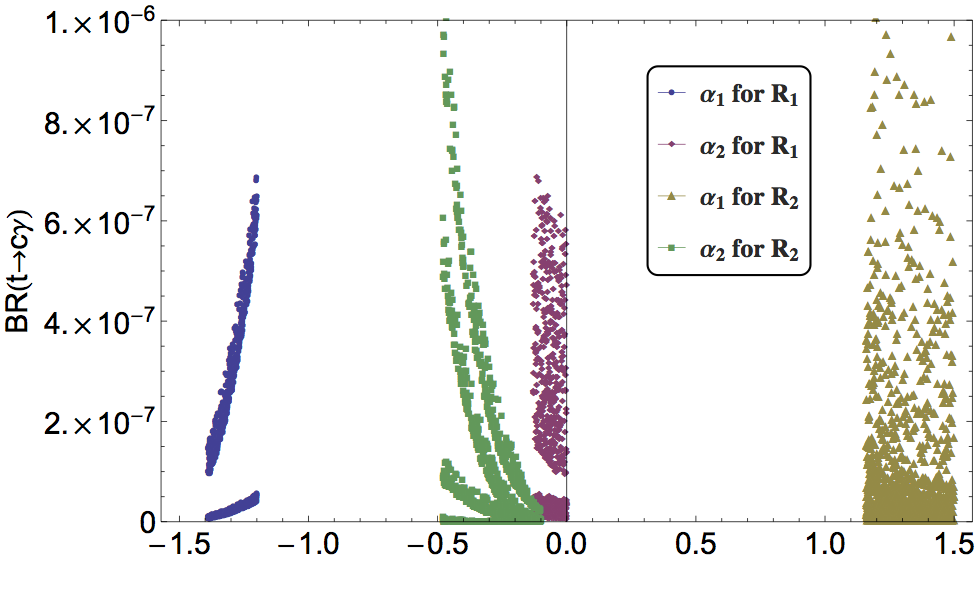}
\caption{The Model III branching ratio for $t\rightarrow c \gamma$ as a function of $\alpha_1$-$\alpha_2$ in regions $R_1$ and $R_2$.}
\label{figure1}
\end{figure}
The following regions can be obtained for $\alpha_1$ and $\alpha_2$ from $0.5\leq R_{\gamma\gamma} \leq 2$ with $m_{H^\pm}=300$ GeV and $\tan\beta=2.5$ \cite{Basso:2012st,Arhrib:2010ju,Krawczyk:2013jta,Chen:2015gaa,Gaitan:2015aia}:
\begin{equation}
R_{1}=\left\{ -1.39\leq \alpha _{1}\leq -1.2\right.\,\textrm{ and}\,\left. -0.13\leq \alpha
_{2}\leq 0\right\},
\end{equation}
and
\begin{equation}
R_{2}=\left\{ 1.16\leq \alpha _{1}\leq 1.5\right. \,\textrm{and}\,\left. -0.48\leq \alpha
_{2}\leq -0.1\right\}.
\end{equation}
The ratio $R_{\gamma\gamma}$ given by 
\begin{equation}
R_{\gamma\gamma}=\frac{\sigma(gg\rightarrow h_1)Br(h_1\rightarrow\gamma\gamma)}{ \sigma(gg\rightarrow h_{SM})Br(h_{SM}\rightarrow\gamma\gamma)},
\end{equation}
allows us to compare the prediction of the THDM with the SM prediction for the Higgs boson diphoton decay.  Fig.(\ref{figure1}) shows $Br(t\rightarrow c\gamma)$ as function of $\alpha_1$ and $\alpha_2$ in the allowed regions $R_1$ and $R_2$ with $\tan\beta=2.5$. The $\textrm{Br}(t\rightarrow c\gamma)$ can be enhanced up to $10^{-6}$ in the regions $R_{1,2}$.
The limits obtained in Model III are less restrictive than those obtained in 2HDM type I and type II, which are of the order $10^{-8}$\cite{Eilam:1990zc,Mele:1998ag}.
In 2021, LHC is expected to reach an integrated luminosity of the order of 300 $\textrm{fb}^{-1}$ \cite{ATLAS:2013hta}. Experiments in LHC Run 3 with this amount of data could find evidence of new physics beyond SM, in particular processes with FCNC. The expected number of events can be naively estimated with the following approximation
\begin{equation}
N\approx \sigma (p\bar{p}\rightarrow t\bar{t}) \textrm{Br}(\bar{t}\rightarrow \bar{b}W)\textrm{Br}(t\rightarrow c\gamma) \mathcal{L}_{int}
\end{equation}
where $\sigma(p\bar{p}\rightarrow t\bar{t})\approx 176 \,pb$ \cite{Agashe:2014kda}, $\mathcal{L}_{int}$ is the integrated luminosity $\sim300\,\textrm{fb}^{-1}$,  $\textrm{Br}(\bar{t}\rightarrow \bar{b}W)\approx 1$ and $Br(t\rightarrow c\gamma)$ is the obtained result in Model III, Eq.(\ref{br}). Due to trigger and selection cuts only a fraction of the produced events are detected by the experiments. An efficiency of 2.4\% is achieved by CMS from simulation of  $tc\gamma$ signal events taking into account all selection criteria~\cite{Khachatryan:2015att}.
Therefore a more realistic estimation of the effective number of events has to be written as $N_{Eff}\approx 0.2\times N $. 

\begin{figure} 
\centering
\includegraphics[scale=0.7]{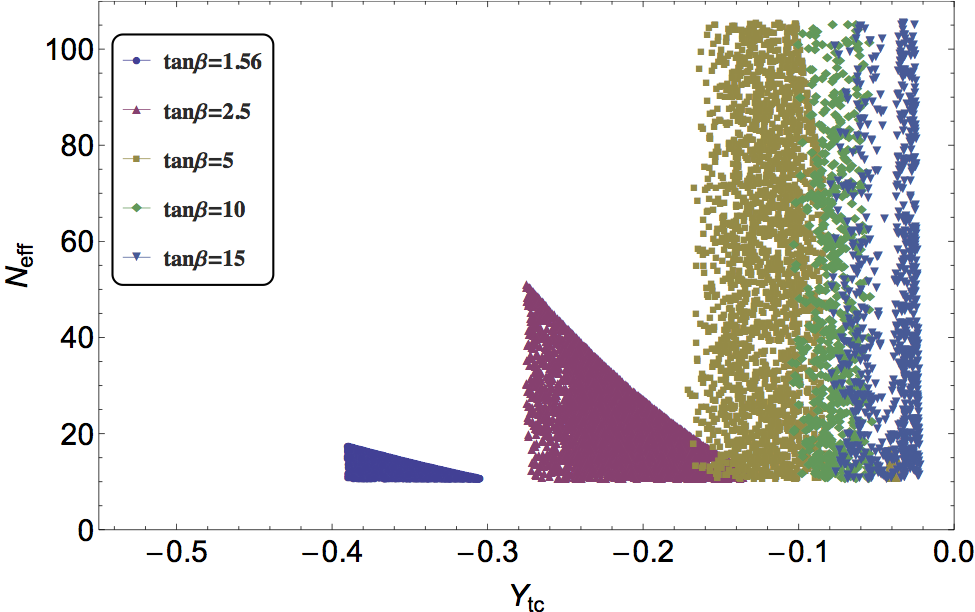}
\caption{Effective number of events for $t\rightarrow c\gamma$ as a function of $Y_{tc}$ for $\tan\beta=1.56,\,2.5,\,5,\,10,\,15$ expected in LHC Run 3.}
\label{Neff_Ytc}
\end{figure}

\begin{figure} 
\centering
\includegraphics[scale=0.7]{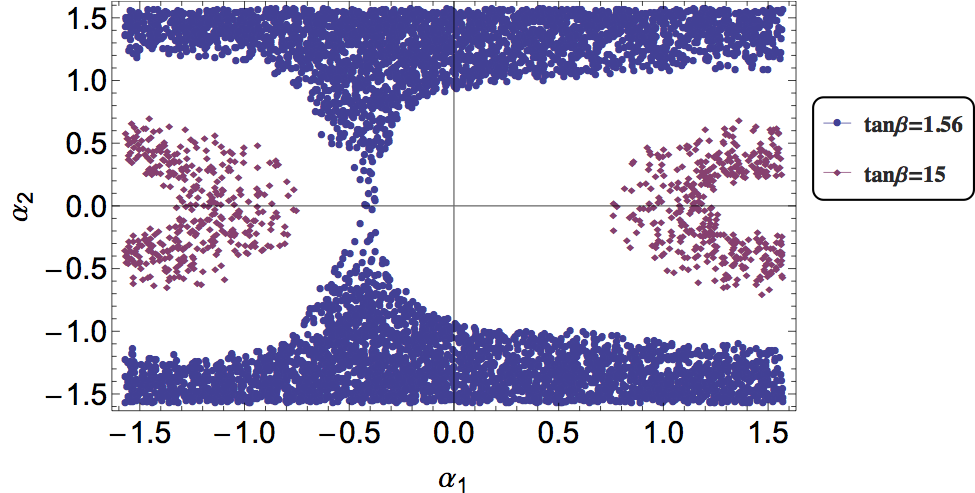}
\caption{Allowed regions for $\alpha_1$ and $\alpha_2$ when $Br(t\rightarrow c \gamma)\sim (1 -10)^{-5}$ is assumed with $-0.385\leqslant Y_{tc} \leqslant -0.307$ for $\tan\beta=1.56$ and $-0.08\leqslant Y_tc \leqslant -0.02$ for $\tan\beta=15$.}
\label{alphas12a}
\end{figure}

\begin{figure} 
\centering
\includegraphics[scale=0.7]{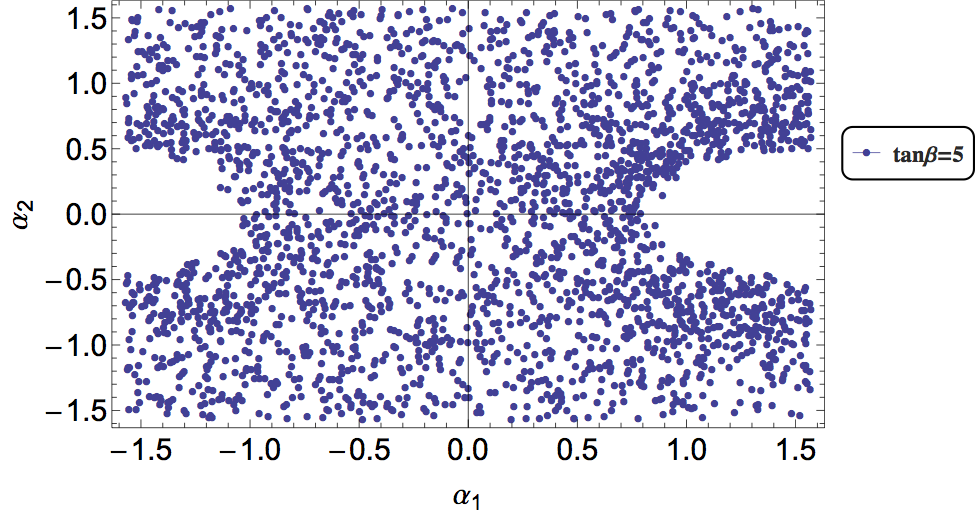}
\caption{Allowed regions for $\alpha_1$ and $\alpha_2$ when $Br(t\rightarrow c \gamma)\sim (1 -10)^{-5}$ is assumed with $-0.173\leqslant Y_{tc} \leqslant -0.035$ for $\tan\beta=5$.}
\label{alphas12b}
\end{figure}

\begin{figure} 
\centering
\includegraphics[scale=0.7]{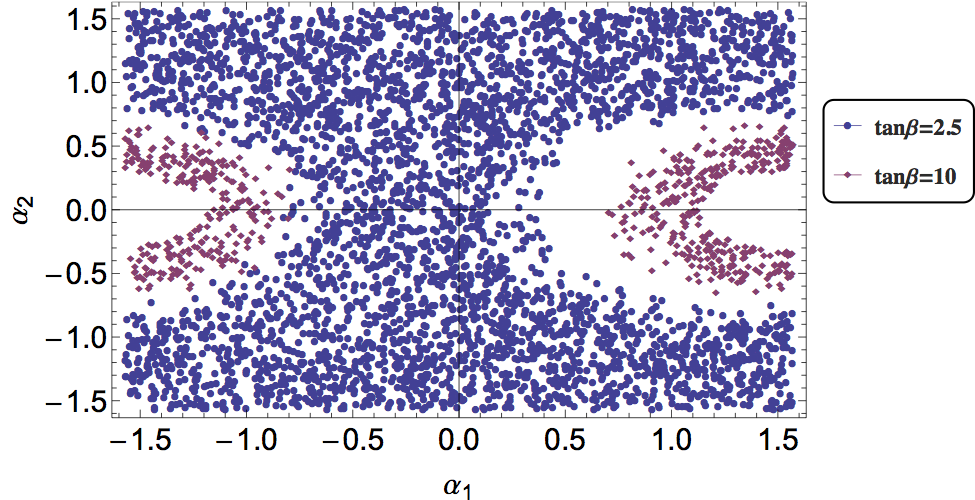}
\caption{Allowed regions for $\alpha_1$ and $\alpha_2$ when $Br(t\rightarrow c \gamma)\sim (1 -10)^{-5}$ is assumed with $-0.267\leqslant Y_{tc} \leqslant -0.135$ for $\tan\beta=2.5$ and $-0.105\leqslant Y_tc \leqslant -0.02$ for $\tan\beta=10$.}
\label{alphas12c}
\end{figure}

\begin{figure} 
\centering
\includegraphics[scale=0.5]{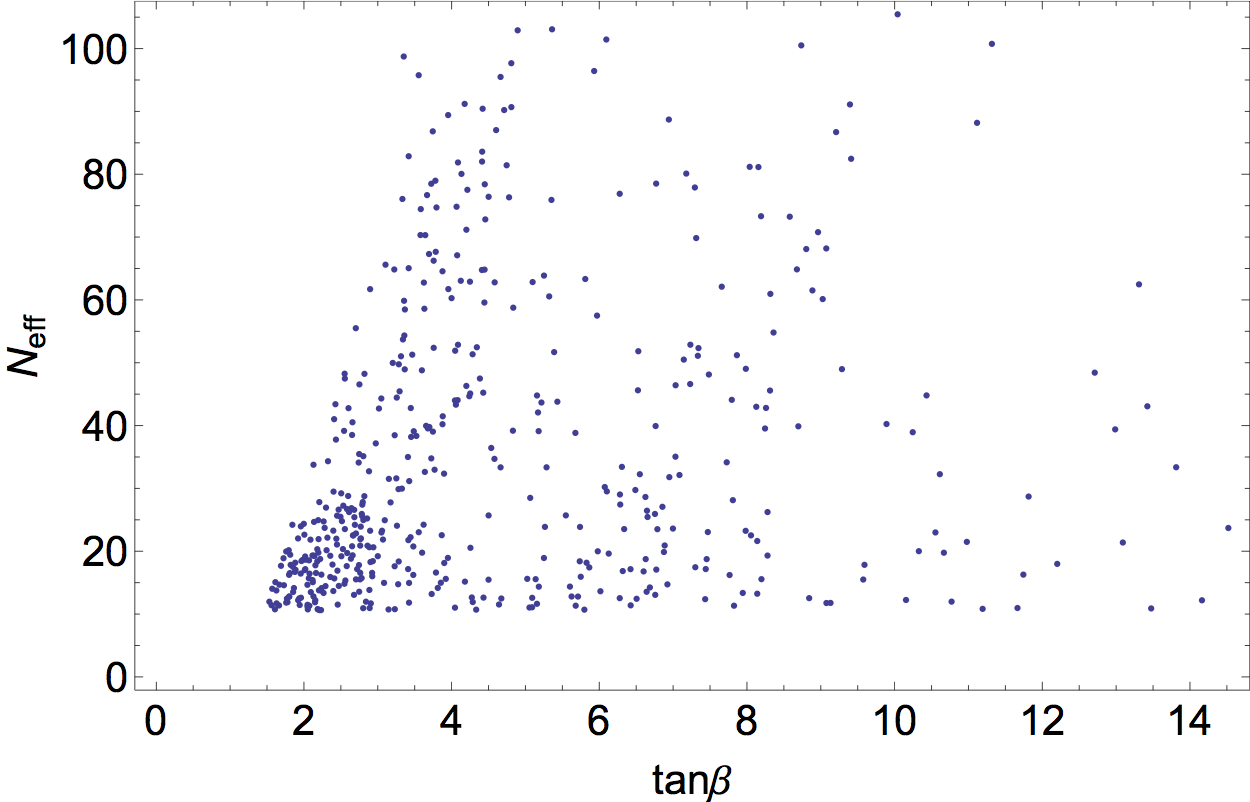}
\caption{Effective number of events for $t\rightarrow c\gamma$ as a function of $\tan\beta$ expected in LHC Run 3.}
\label{Neff_tanbeta}
\end{figure}

The limit that is expected to be reached in future experiments is $\textrm{Br}(t\rightarrow c\gamma)\sim 10^{-5}$~\cite{ATLAS:2013hta}. If we consider this expected limit as $\textrm{Br}(t\rightarrow c\gamma)\sim (1 -10)^{-5}$ with $N_{Eff}\geqslant 1$ and impose the  restrictions discussed in the previous section, then the $N_{Eff}$ can be estimated for fixed values of $\tan\beta$. Fig.(\ref{Neff_Ytc}) shows $N_{Eff}$ as a function of $Y_{tc}$. The mixing parameters $\alpha_{1,2}$ are also bounded by same constraints and the allowed values of the $\alpha_{1,2}$ are shown in Fig.(\ref{alphas12a}), Fig.(\ref{alphas12b}) and Fig.(\ref{alphas12c}) for fixed $\tan\beta$. The numerical values for $\tan\beta$ are fixed by the representative values $\tan\beta=1.56,\,2.5,\,5,\,10,\,15$; however, the $N_{Eff}$ as function of $\tan\beta$ with the above restrictions is shown in Fig.(\ref{Neff_tanbeta}). We find that there is more than one event, $N_{Eff}\geqslant 1$, from $\tan\beta \geqslant 1.56$.
%
%
%
%
%
%
%
%
%
%
%
%
%
%
%
%
%
\section{Conclusions}
\label{sec6}
The expression for rare top decay $t\rightarrow c \gamma$ was calculated at one loop due to the FCNSI in an extended model with two scalar doublets. The SM predicted value for the $Br(t\rightarrow c \gamma)$ is extremely suppressed from LHC sensitivity, while in the considered THDM type III with mixing in the neutral scalars the same branching ratio has been increased making it possible to test rare decays in future experiments. In this work we have studied a theoretical framework where  $\textrm{Br}(t\rightarrow c \gamma)\sim 10^{-5}$ can be viable for specific values of mixing parameters. 

If the $t\rightarrow c \gamma$ decay is observed in LHC, it will provide an important evidence of physics beyond SM.
With the allowed regions for the $\alpha_1$, $\alpha_2$ and $\tan\beta\simeq 2.5$, Model III predicts $\textrm{Br}(t\rightarrow c \gamma)\sim10^{-6}$. Model III, with an integrated luminosity of $300\,\textrm{fb}^{-1}$, predicts up to $N_{Eff}\approx100$ events for $t\rightarrow c \gamma$ decay with $\alpha_1$, $\alpha_2$ and $\tan\beta$ given in previous section.
\section*{Acknowledgments}
This work was supported by projects PAPIIT-IN113916 in DGAPA-UNAM, PIAPIVC07 in FES-Cuautitlan UNAM and \emph{Sistema Nacional de Investigadores} (SNI) M\'exico. R. Martinez thanks COLCIENCIAS for the financial support. E. A. Garc\'es thanks CONACYT postdoctoral grant.
Authors thank L. D\'iaz-Cruz for useful discussions.

\begin{thebibliography}{9}
%
\bibitem{Aad:2012tfa} 
  G.~Aad {\it et al.} [ATLAS Collaboration],
  Phys.\ Lett.\ B {\bf 716}, 1 (2012)
  doi:10.1016/j.physletb.2012.08.020
  [arXiv:1207.7214 [hep-ex]].

\bibitem{Chatrchyan:2012xdj} 
  S.~Chatrchyan {\it et al.} [CMS Collaboration],
  Phys.\ Lett.\ B {\bf 716}, 30 (2012)
  doi:10.1016/j.physletb.2012.08.021
  [arXiv:1207.7235 [hep-ex]].

\bibitem{hunter}
J. F. Gunion, H.E. Haber, G.L.Kane, and S. Dawson, The Higgs Hunters Guide Westview Press, Boulder, CO, (2000)

\bibitem{Haber:1978jt} 
  H.~E.~Haber, G.~L.~Kane and T.~Sterling,
  Nucl.\ Phys.\ B {\bf 161}, 493 (1979).
  doi:10.1016/0550-3213(79)90225-6

\bibitem{Hall:1981bc} 
  L.~J.~Hall and M.~B.~Wise,
  Nucl.\ Phys.\ B {\bf 187}, 397 (1981).
  doi:10.1016/0550-3213(81)90469-7

\bibitem{Donoghue:1978cj} 
  J.~F.~Donoghue and L.~F.~Li,
  Phys.\ Rev.\ D {\bf 19}, 945 (1979).
  doi:10.1103/PhysRevD.19.945

\bibitem{Atwood:1996vj} 
  D.~Atwood, L.~ and A.~Soni,
  Phys.\ Rev.\ D {\bf 55}, 3156 (1997)
  doi:10.1103/PhysRevD.55.3156
  [hep-ph/9609279].

\bibitem{Barroso:2012wz} 
  A.~Barroso, P.~M.~Ferreira, R.~Santos and J.~P.~Silva,
  Phys.\ Rev.\ D {\bf 86}, 015022 (2012)
  doi:10.1103/PhysRevD.86.015022
  [arXiv:1205.4247 [hep-ph]].
\bibitem{Lavoura:1994fv} 
  L.~Lavoura and J.~P.~Silva,
  Phys.\ Rev.\ D {\bf 50}, 4619 (1994)
  doi:10.1103/PhysRevD.50.4619
  [hep-ph/9404276].
\bibitem{Fontes:2015mea} 
  D.~Fontes, J.~C.~Romão, R.~Santos and J.~P.~Silva,
  JHEP {\bf 1506}, 060 (2015)
  doi:10.1007/JHEP06(2015)060
  [arXiv:1502.01720 [hep-ph]].

\bibitem{ATLAS:2013nma} 
  [ATLAS Collaboration],
  ATLAS-CONF-2013-013.
\bibitem{Khachatryan:2014kca} 
  V.~Khachatryan {\it et al.} [CMS Collaboration],
  Phys.\ Rev.\ D {\bf 92}, no. 1, 012004 (2015)
  doi:10.1103/PhysRevD.92.012004
  [arXiv:1411.3441 [hep-ex]].


\bibitem{Cheng:1987rs} 
  T.~P.~Cheng and M.~Sher,
  Phys.\ Rev.\ D {\bf 35}, 3484 (1987).
  doi:10.1103/PhysRevD.35.3484

\bibitem{Crivellin:2013wna} 
  A.~Crivellin, A.~Kokulu and C.~Greub,
  Phys.\ Rev.\ D {\bf 87}, no. 9, 094031 (2013)
  doi:10.1103/PhysRevD.87.094031
  [arXiv:1303.5877 [hep-ph]].

\bibitem{Fritzsch:1977za} 
  H.~Fritzsch,
  Phys.\ Lett.\ B {\bf 70}, 436 (1977).
  doi:10.1016/0370-2693(77)90408-7
\bibitem{Fritzsch:1977vd} 
  H.~Fritzsch,
  Phys.\ Lett.\ B {\bf 73}, 317 (1978).
  doi:10.1016/0370-2693(78)90524-5
\bibitem{Fritzsch:1979zq} 
  H.~Fritzsch,
  Nucl.\ Phys.\ B {\bf 155}, 189 (1979).
  doi:10.1016/0550-3213(79)90362-6


\bibitem{Inoue:2014} 
Satoru Inoue,Michael J. Ramsey-Musolf and Yue Zhang {\it Phys. Rev. D}{\bf 89},115023 (2014); 
Joachim Brod, Ulrich Haisch and Jure Zupan {\it JHEP} 11(2013)180.

\bibitem{Carcamo:2006dp} 
  A.~E.~Carcamo Hernandez, R.~Martinez and J.~A.~Rodriguez,
  Eur.\ Phys.\ J.\ C {\bf 50}, 935 (2007)
  doi:10.1140/epjc/s10052-007-0264-0
  [hep-ph/0606190].
\bibitem{Cvetic:1997zd} 
  G.~Cvetic, S.~S.~Hwang and C.~S.~Kim,
  Int.\ J.\ Mod.\ Phys.\ A {\bf 14}, 769 (1999)
  doi:10.1142/S0217751X99000385
  [hep-ph/9706323].
\bibitem{Cvetic:1998uw} 
  G.~Cvetic, C.~S.~Kim and S.~S.~Hwang,
  Phys.\ Rev.\ D {\bf 58}, 116003 (1998)
  doi:10.1103/PhysRevD.58.116003
  [hep-ph/9806282].

\bibitem{Shu:2013uua} 
  J.~Shu and Y.~Zhang,
  Phys.\ Rev.\ Lett.\  {\bf 111}, no. 9, 091801 (2013)
  doi:10.1103/PhysRevLett.111.091801
  [arXiv:1304.0773 [hep-ph]].
\bibitem{Morrissey:2012db} 
  D.~E.~Morrissey and M.~J.~Ramsey-Musolf,
  New J.\ Phys.\  {\bf 14}, 125003 (2012)
  doi:10.1088/1367-2630/14/12/125003
  [arXiv:1206.2942 [hep-ph]].

\bibitem{AguilarSaavedra:2004wm} 
  J.~A.~Aguilar-Saavedra,
  Acta Phys.\ Polon.\ B {\bf 35}, 2695 (2004)
  [hep-ph/0409342].
\bibitem{Grzadkowski:1990sm} 
  B.~Grzadkowski, J.~F.~Gunion and P.~Krawczyk,
  Phys.\ Lett.\ B {\bf 268}, 106 (1991).
  doi:10.1016/0370-2693(91)90931-F
\bibitem{Mele:1999zx} 
  B.~Mele,
  hep-ph/0003064.
\bibitem{Mele:1998ag} 
  B.~Mele, S.~Petrarca and A.~Soddu,
  Phys.\ Lett.\ B {\bf 435}, 401 (1998)
  doi:10.1016/S0370-2693(98)00822-3
  [hep-ph/9805498].
\bibitem{Gabrielli:2011yn} 
  E.~Gabrielli and B.~Mele,
  Phys.\ Rev.\ D {\bf 83}, 073009 (2011)
  doi:10.1103/PhysRevD.83.073009
  [arXiv:1102.3361 [hep-ph]].
\bibitem{Eilam:1990zc} 
  G.~Eilam, J.~L.~Hewett and A.~Soni,
  Phys.\ Rev.\ D {\bf 44}, 1473 (1991)
  Erratum: [Phys.\ Rev.\ D {\bf 59}, 039901 (1999)].
  doi:10.1103/PhysRevD.44.1473, 10.1103/PhysRevD.59.039901
\bibitem{AguilarSaavedra:2002ns} 
  J.~A.~Aguilar-Saavedra and B.~M.~Nobre,
  Phys.\ Lett.\ B {\bf 553}, 251 (2003)
  doi:10.1016/S0370-2693(02)03230-6
  [hep-ph/0210360].
\bibitem{Larios:2006pb} 
  F.~Larios, R.~Martinez and M.~A.~Perez,
  Int.\ J.\ Mod.\ Phys.\ A {\bf 21}, 3473 (2006)
  doi:10.1142/S0217751X06033039
  [hep-ph/0605003].

\bibitem{DiazCruz:1989ub} 
  J.~L.~Diaz-Cruz, R.~Martinez, M.~A.~Perez and A.~Rosado,
  Phys.\ Rev.\ D {\bf 41}, 891 (1990).
  doi:10.1103/PhysRevD.41.891

\bibitem{Agashe:2014kda} 
  K.~A.~Olive {\it et al.} [Particle Data Group Collaboration],
  Chin.\ Phys.\ C {\bf 38}, 090001 (2014).
  doi:10.1088/1674-1137/38/9/090001

\bibitem{Luke:1993cy} 
  M.~E.~Luke and M.~J.~Savage,
  Phys.\ Lett.\ B {\bf 307}, 387 (1993)
  doi:10.1016/0370-2693(93)90238-D
  [hep-ph/9303249].

\bibitem{Atwood:1995ud}
  D.~Atwood, L.~ and A.~Soni,
  Phys.\ Rev.\ D {\bf 53}, 1199 (1996)
  doi:10.1103/PhysRevD.53.1199
  [hep-ph/9506243].
\bibitem{Atwood:1996vw} 
  D.~Atwood, L.~ and A.~Soni,
  Phys.\ Rev.\ D {\bf 54}, 3296 (1996)
  doi:10.1103/PhysRevD.54.3296
  [hep-ph/9603210].
\bibitem{Atwood:1995ej} 
  D.~Atwood, L.~ and A.~Soni,
  Phys.\ Rev.\ Lett.\  {\bf 75}, 3800 (1995)
  doi:10.1103/PhysRevLett.75.3800
  [hep-ph/9507416].

\bibitem{Arhrib:2005nx} 
  A.~Arhrib,
  Phys.\ Rev.\ D {\bf 72}, 075016 (2005)
  doi:10.1103/PhysRevD.72.075016
  [hep-ph/0510107].

\bibitem{Branco:2011iw} 
  G.~C.~Branco, P.~M.~Ferreira, L.~Lavoura, M.~N.~Rebelo, M.~Sher and J.~P.~Silva,
  Phys.\ Rept.\  {\bf 516}, 1 (2012)
  doi:10.1016/j.physrep.2012.02.002
  [arXiv:1106.0034 [hep-ph]].
\bibitem{AguilarSaavedra:2002kr} 
  J.~A.~Aguilar-Saavedra,
  Phys.\ Rev.\ D {\bf 67}, 035003 (2003)
  Erratum: [Phys.\ Rev.\ D {\bf 69}, 099901 (2004)]
  doi:10.1103/PhysRevD.69.099901, 10.1103/PhysRevD.67.035003
  [hep-ph/0210112].

\bibitem{Han:2009zm} 
  X.~F.~Han, L.~Wang and J.~M.~Yang,
  Phys.\ Rev.\ D {\bf 80}, 015018 (2009)
  doi:10.1103/PhysRevD.80.015018
  [arXiv:0903.5491 [hep-ph]].
\bibitem{HongSheng:2007ve} 
  H.~Hong-Sheng,
  Phys.\ Rev.\ D {\bf 75}, 094010 (2007)
  doi:10.1103/PhysRevD.75.094010
  [hep-ph/0703067 [HEP-PH]].
\bibitem{GonzalezSprinberg:2007zz} 
  G.~A.~Gonzalez-Sprinberg, R.~Martinez and J.~A.~Rodriguez,
  Eur.\ Phys.\ J.\ C {\bf 51}, 919 (2007).
  doi:10.1140/epjc/s10052-007-0344-1
\bibitem{Cao:2007dk} 
  J.~J.~Cao, G.~Eilam, M.~Frank, K.~Hikasa, G.~L.~Liu, I.~Turan and J.~M.~Yang,
  Phys.\ Rev.\ D {\bf 75}, 075021 (2007)
  doi:10.1103/PhysRevD.75.075021
  [hep-ph/0702264].
\bibitem{Han:2003qe} 
  T.~Han, K.~i.~Hikasa, J.~M.~Yang and X.~m.~Zhang,
  Phys.\ Rev.\ D {\bf 70}, 055001 (2004)
  doi:10.1103/PhysRevD.70.055001
  [hep-ph/0312129].

\bibitem{Diaz:2001vj} 
  R.~A.~Diaz, R.~Martinez and J.~Alexis Rodriguez,
  hep-ph/0103307.
\bibitem{Gaitan-Lozano:2014nka} 
  R.~Gaitan-Lozano, R.~Martinez and J.~H.~M.~de Oca,
  arXiv:1407.3318 [hep-ph].

\bibitem{Dedes:2014asa} 
  A.~Dedes, M.~Paraskevas, J.~Rosiek, K.~Suxho and K.~Tamvakis,
  JHEP {\bf 1411}, 137 (2014)
  doi:10.1007/JHEP11(2014)137
  [arXiv:1409.6546 [hep-ph]].

\bibitem{Accomando:2006ga} 
  E.~Accomando {\it et al.},
  hep-ph/0608079.
\bibitem{Niezurawski:2004ui} 
  P.~Niezurawski, A.~F.~Zarnecki and M.~Krawczyk,
  JHEP {\bf 0502}, 041 (2005)
  doi:10.1088/1126-6708/2005/02/041
  [hep-ph/0403138].

\bibitem{Haber:1993an} 
  H.~E.~Haber and R.~Hempfling,
  Phys.\ Rev.\ D {\bf 48}, 4280 (1993)
  doi:10.1103/PhysRevD.48.4280
  [hep-ph/9307201].

\bibitem{Ginzburg:2004vp} 
  I.~F.~Ginzburg and M.~Krawczyk,
  Phys.\ Rev.\ D {\bf 72}, 115013 (2005)
  doi:10.1103/PhysRevD.72.115013
  [hep-ph/0408011].

\bibitem{ElKaffas:2006gdt} 
  A.~W.~El Kaffas, W.~Khater, O.~M.~Ogreid and P.~Osland,
  Nucl.\ Phys.\ B {\bf 775}, 45 (2007)
  doi:10.1016/j.nuclphysb.2007.03.041
  [hep-ph/0605142].

\bibitem{Basso:2012st} 
  L.~Basso, A.~Lipniacka, F.~Mahmoudi, S.~Moretti, P.~Osland, G.~M.~Pruna and M.~Purmohammadi,
  JHEP {\bf 1211}, 011 (2012)
  doi:10.1007/JHEP11(2012)011
  [arXiv:1205.6569 [hep-ph]].
\bibitem{Arhrib:2010ju} 
  A.~Arhrib, E.~Christova, H.~Eberl and E.~Ginina,
  JHEP {\bf 1104}, 089 (2011)
  doi:10.1007/JHEP04(2011)089
  [arXiv:1011.6560 [hep-ph]].

\bibitem{Krawczyk:2013jta} 
  M.~Krawczyk, D.~Sokolowska, P.~Swaczyna and B.~Swiezewska,
  JHEP {\bf 1309}, 055 (2013)
  doi:10.1007/JHEP09(2013)055
  [arXiv:1305.6266 [hep-ph]].
\bibitem{Chen:2015gaa} 
  C.~Y.~Chen, S.~Dawson and Y.~Zhang,
  JHEP {\bf 1506}, 056 (2015)
  doi:10.1007/JHEP06(2015)056
  [arXiv:1503.01114 [hep-ph]].

\bibitem{Degrassi:2000qf} 
  G.~Degrassi, P.~Gambino and G.~F.~Giudice,
  JHEP {\bf 0012}, 009 (2000)
  doi:10.1088/1126-6708/2000/12/009
  [hep-ph/0009337].
\bibitem{Misiak:2006zs} 
  M.~Misiak {\it et al.},
  Phys.\ Rev.\ Lett.\  {\bf 98}, 022002 (2007)
  doi:10.1103/PhysRevLett.98.022002
  [hep-ph/0609232].
\bibitem{Lunghi:2006hc} 
  E.~Lunghi and J.~Matias,
  JHEP {\bf 0704}, 058 (2007)
  doi:10.1088/1126-6708/2007/04/058
  [hep-ph/0612166].
\bibitem{Gomez:2006uv} 
  M.~E.~Gomez, T.~Ibrahim, P.~Nath and S.~Skadhauge,
  Phys.\ Rev.\ D {\bf 74}, 015015 (2006)
  doi:10.1103/PhysRevD.74.015015
  [hep-ph/0601163].
\bibitem{Barenboim:2013bla} 
  G.~Barenboim, C.~Bosch, M.~L.~López-Ibañez and O.~Vives,
  JHEP {\bf 1311}, 051 (2013)
  doi:10.1007/JHEP11(2013)051
  [arXiv:1307.5973 [hep-ph]].

\bibitem{Chen:2001fja} 
  S.~Chen {\it et al.} [CLEO Collaboration],
  Phys.\ Rev.\ Lett.\  {\bf 87}, 251807 (2001)
  doi:10.1103/PhysRevLett.87.251807
  [hep-ex/0108032].
\bibitem{Abe:2001hk} 
  K.~Abe {\it et al.} [Belle Collaboration],
  Phys.\ Lett.\ B {\bf 511}, 151 (2001)
  doi:10.1016/S0370-2693(01)00626-8
  [hep-ex/0103042].
\bibitem{Lees:2012wg} 
  J.~P.~Lees {\it et al.} [BaBar Collaboration],
  Phys.\ Rev.\ D {\bf 86}, 052012 (2012)
  doi:10.1103/PhysRevD.86.052012
  [arXiv:1207.2520 [hep-ex]].
\bibitem{Lees:2012ufa} 
  J.~P.~Lees {\it et al.} [BaBar Collaboration],
  Phys.\ Rev.\ D {\bf 86}, 112008 (2012)
  doi:10.1103/PhysRevD.86.112008
  [arXiv:1207.5772 [hep-ex]].
\bibitem{Aubert:2007my} 
  B.~Aubert {\it et al.} [BaBar Collaboration],
  Phys.\ Rev.\ D {\bf 77}, 051103 (2008)
  doi:10.1103/PhysRevD.77.051103
  [arXiv:0711.4889 [hep-ex]].
  
%
\bibitem{Amhis:2014hma}
  Y.~Amhis {\it et al.}  [Heavy Flavor Averaging Group (HFAG) Collaboration],
  arXiv:1412.7515 [hep-ex].

\bibitem{Glashow:1976nt}
  S.~L.~Glashow and S.~Weinberg,
  Phys.\ Rev.\ D {\bf 15}, 1958 (1977).

\bibitem{euro2014}
\bibitem{Gaitan:2013yfa} 
  R.~Gait\'an, R.~Martinez, J.~H.~Montes de Oca and S.~R.~Romo,
  Eur.\ Phys.\ J.\ C {\bf 74}, no. 3, 2788 (2014)
  doi:10.1140/epjc/s10052-014-2788-4
  [arXiv:1312.0044 [hep-ph]].

\bibitem{Gaitan:2015aia} 
  R.~Gaitan, E.~A.~Garces, J.~H.~M.~de Oca and R.~Martinez,
  Phys.\ Rev.\ D {\bf 92}, no. 9, 094025 (2015)
  doi:10.1103/PhysRevD.92.094025
  [arXiv:1505.04168 [hep-ph]].

\bibitem{Khachatryan:2015att} 
  V.~Khachatryan {\it et al.} [CMS Collaboration],
  JHEP {\bf 1604}, 035 (2016)
  doi:10.1007/JHEP04(2016)035
  [arXiv:1511.03951 [hep-ex]].

\bibitem{ATLAS:2013hta}
  [ATLAS Collaboration],
  arXiv:1307.7292 [hep-ex].
\end{thebibliography}
\end{document}